\documentclass[proof]{WileyASNA-v1}

\usepackage{slashed}
\usepackage{amsmath}
\newcommand{\punto}{\!\cdot\!}
\newcommand{\bs}[1]{{\boldsymbol #1}}

\articletype{}%

\received{}
\revised{ }
\accepted{ }

\begin{document}

\title{Pions near condensation under compact star conditions
}

\author[1]{Cristi\'an Villavicencio*}

\author[2]{Marcelo Loewe}

\author[3]{Alfredo Raya}

\authormark{Villavicencio \textsc{et al}}

\address[1]{\orgdiv{Departamento de Ciencias B\'asicas - Facultad de Ciencias}, \orgname{Universidad del B\'io-B\'io}, \orgaddress{ \city{Chill\'an}, \country{Chile}}}

\address[2]{\orgdiv{Instituto de F\'isica}, \orgname{Pontificia Universidad Cat\'olica de Chile}, \orgaddress{\city{Santiago}, \country{Chile}}}

\address[3]{\orgdiv{Instituto de F\'sica y Matem\'aticas}, \orgname{Universidad Michoacana de San Nicol\'as de Hidalgo}, \orgaddress{\city{Morelia}, \state{Michoac\'an}, \country{M\'exico}}}

\corres{*Avda. Andr\'es Bello 720, Casilla 447 - CP: 3800708, Chill\'an, Chile. \\
\email{cvillavi@uc.cl}}


\abstract{The behavior of pions is studied in systems where their normal leptonic decay is forbidden. When thermal fluctuations are present, a low decay rate is generated, and as a consequence of lepton recombination, the amount of pions remains almost unaltered. Compact stars conditions are favorable for the formation of such intermediate state of charged pions: near condensation and almost stable, leading to a continuum source of anti-neutrinos. In particular, protoneutron stars could be an scenario where this state of matter is relevant.
}




\maketitle
%
%

\section{Introduction}

Motivated by the possibility of new hadronic phases under extreme conditions, interesting phenomena can occur in dense media. 
In particular, charged pions in a degenerated leptonic system, as in the case in neutron stars, can experience interesting changes in their behavior. 
 
The influence of pion or kaon condensate in the decay rate of neutron stars is still an open question.
In particular, the formation of a pion Bose-Einstein condensation in chemical equilibrium with leptons needs that the electron chemical potential reaches the pion mass. 
However, it is not clear if repulsive $s$-wave or attractive $p$-wave dominates, with the consequences that in the first case pion mass increases, and in the second case pion mass diminishes \cite{Ohnishi:2008ng}. 
Also the existence of hyperons reduces the electron chemical potential \cite{Heiselberg:1999mq}. 
  
The population of pions in the normal phase depends strongly on temperature, so, for cold neutron stars, their influence is negligible.
However, for protoneutron stars the amount of pions in normal phase can increase considerably. 
Nevertheless, because of their fast decayment, pions in normal phase are usually ignored in the literature and rarely considered in the description of relative population of particles inside neutron stars or protoneutron stars \cite{Nakazato:2008su}.

Charged pions decay mainly into muons (the 99.99\% of them) and eventually into electrons (< 0.01\%).
Because of Pauli blocking, if the system is degenerated with respects of leptons, the leptonic Fermi levels are occupied, forbidding the pion decay since there is no room in the phase space to locate the emerging lepton . 
This scenario provides an interesting question about the role of pions in terms of stability and the emissivity of neutrinos when thermal effects are significant.

Here we will explore the possibility of finding charged pions in this low decay process, which we denote it conveniently pions in {\it metastable} state, in the language  of nuclear physics for denoting long-lived isotopes.
For a more detailed description of this work, see \citep{Loewe:2016wsk} and also references therein. 
Here we will present the main ideas and results.

\section{Pion decay rate}
\label{decay_width}

The decay rate of pions into leptons is originated from electroweak process, where the Lagrangian term that governs this interaction is given by 
\begin{equation}
{\cal L}_{\pi\ell} = 
f_\pi G_F[ 
\bar\psi_{\nu_\ell} \slashed{D}\pi^+(1-\gamma_5)\psi_{\ell}
+\bar\psi_{\ell}\slashed{D}\pi^-(1-\gamma_5)\psi_{\nu_\ell}],
\label{L_pi-l}
\end{equation}
where $\pi^\pm$ are the charged pion fields, $\psi_\ell$ describes the lepton field (muons or electrons) and $\psi_{\nu_\ell}$ the associated neutrino field. 
The derivative 
\begin{equation}
D_\alpha\pi^\pm=(\partial_\alpha\pm i\mu_\pi \delta_{\alpha 0})\pi^\pm
\end{equation}
includes the pion chemical potential, and $G_F=1.17\times 10^{-5}$~GeV$^{-2}$ is the Fermi constant.

The respective free Lagrangian for the fields involved are
\begin{align}
{\cal L}_{\pi} &= D_\alpha\pi^+ D^\alpha\pi^- - m_\pi^2\pi^+\pi^-,\\
{\cal L}_\ell &= \bar\psi_\ell [i\slashed{\partial}+\mu_\ell\gamma_0-m_\ell]\psi_\ell\\
{\cal L}_{\nu_\ell} &= \bar\psi_{\nu_\ell}[i\slashed{\partial}+\mu_{\nu_\ell}\gamma_0]\psi_{\nu_\ell}.
\label{L_f}
\end{align}
where their associated chemical potentials are included. 
This is an effective model in the sense that we consider all the hadronic parameters as density dependent. 
This affects directly the pion mass and pion decay constant .

The decay rate can be obtained from the retarded $\pi^-$ propagator
\begin{equation}
D_{\pi^-}^{\mathrm{ret}}(p) =\left.\frac{i}{(p_0+ \mu_\pi)^2-E_\pi^2-\Pi(p)}\right|_{p_0\to p_0+i\epsilon}.
\end{equation}
The denominator of the dressed propagator is expanded around the physical real pole, obtaining a Breit-Wigner propagator with a momentum dependent decay rate defined as
\begin{equation}
\Gamma_{\pi^-} = -\frac{1}{E_\pi}\mathrm{Im} ~\Pi(E_\pi-\mu_\pi+i\epsilon,~{\boldsymbol p}),
\label{def-Gamma}
\end{equation}
where $\Pi(p_0,\bs{p})$ is the charged pion self-energy and where $E_\pi=\sqrt{\bs{p}^2+m_\pi^2}$  is the pion energy. 
The imaginary part of the pion comes from the one-loop correction diagram from the weak interaction term described in Eq.(\ref{L_pi-l}) and thermal effects are considered using the Matsubara formalism.

We consider that the leptons involved are in chemical equilibrium from the process $\mu^- \to e+\bar\nu_e+\nu_\mu$, therefore their chemical potentials follow the relation $\mu_\mu -\mu_{\nu_\mu} = \mu_e -\mu_{\nu_e} \equiv \mu_\ell -\mu_{\nu_\ell}$. 
Also negative charged pions will be in chemical equilibrium from the process $\pi^-\to \ell +\bar\nu_\ell$, so pion chemical potential follows the relation $\mu_{\pi}=  \mu_\ell -\mu_{\nu_\ell}$. 
Under such considerations, the decay rate including thermal effects is
\begin{align}
\Gamma_{\pi^-}(\bs{p}) & = \bar\Gamma_{\pi\ell} \frac{m_\pi}{E_\pi}
\left[1+\frac{T}{2a_\ell |{\boldsymbol p}|}\ln\left(
\frac{1+e^{-(E_\ell^+ - \mu_\ell)/T}}{1+e^{-(E_\ell^- - \mu_\ell)/T}}
\right)\right.
\nonumber\\&\qquad\quad\quad\;
\left.+\frac{T}{2a_\ell |{\boldsymbol p}|}\ln\left(
\frac{1+e^{-(E_{\nu_\ell}^+ + \mu_{\nu_\ell})/T}}{1+e^{-(E_{ \nu_\ell}^- + \mu_{\nu_\ell})/T}}
\right)\right]\;,
\label{Gamma_pi}
\end{align}
were $\bar\Gamma_{\pi\ell}$ is the pion decay width at zero temperature and chemical potential, the lepton and neutrino energy terms are defined as $E_\ell^\pm  =  (1-a_\ell)E_\pi \pm a_\ell |{\boldsymbol p}|$, and $E_{\nu_\ell}^\pm   = a_\ell( E_\pi \pm  |{\boldsymbol p}|)$,  and with $a_\ell = (m_\pi^2-m_\ell^2)/2m_\pi^2$.
It is assumed, of course, that $m_\pi>m_\ell$.

A more enlightening expression can be obtained in the rest frame:
\begin{equation}
\Gamma_{\pi^-}(0) =  \bar \Gamma_{\pi\ell}
\left[1-n_F( M_\ell-\mu_\ell) - n_F( M_{\nu_\ell} +\mu_{\nu_\ell})\right],
\label{Gamma1}
\end{equation}
where the mass terms are defined as $M_\ell = (m_\pi^2+m_\ell^2)/2m_\pi$, and $M_{\nu_\ell} = (m_\pi^2-m_\ell^2)/2m_\pi$. From this expression one can see that the thermal corrections can be expressed as $1-n_\ell - n_{\bar\nu_\ell} =(1-n_\ell)(1-n_{\bar\nu_\ell})-n_\ell n_{\bar\nu_\ell}$, which is interpreted as the availability to create a lepton and an antineutrino in the thermal bath minus the probability of finding a lepton and an antineutrino in the thermal bath. 
This is often referred as the direct and inverse process $\Gamma_d - \Gamma_i$. 
The macroscopic interpretation of the decay rate obtained in this way is explained in \citep{Weldon:1983jn}:  the rate at which the slightly out-of-equilibrium system can reach thermal equilibrium.
Although it is not well understood, this interpretation is the most accepted one.

A direct consequence in Eq. (\ref{Gamma1}) can be observed at zero temperature. 
The decay rate vanishes if $\mu_\ell > M_\ell$. 
This is the phenomenon we want to study in more detail.

\section{Metastability conditions}

The effect of Pauli blocking inhibits the the particle decay, and only thermal fluctuations can overcome this situation. 
We can write the decay rate separated as the contribution at zero temperature and the thermal fluctuations
\begin{equation}
\Gamma_{\pi^-} = \Gamma_{0} + \delta\Gamma_T,
\end{equation}
where $\Gamma_0$ is the decay rate in the $T=0$ limit and $\delta\Gamma_T$ the thermal contribution.
We will denote the pion system as metastable if $\Gamma_0 = 0$. 
This definition, as was pointed in the introduction, is in analogy of nuclear physics definition of particles with finite lifetime but considerably longer than usual.
From Eq. (\ref{Gamma_pi}), we have
\begin{align}
\Gamma_0 = \bar\Gamma_{\pi\ell}\frac{m_\pi}{E_\pi}\Bigg[ 1 
&+ \frac{\mu_\ell-E_\ell^+}{2a_\ell |\bs{p}|}\theta(\mu_\ell-E_\ell^+)
\nonumber \\ &
-\frac{\mu_\ell-E_\ell^-}{2a_\ell |\bs{p}|}\theta(\mu_\ell-E_\ell^-)
\Bigg].
\end{align}
From the above equation is easy to see that $\Gamma_0$ vanishes for $\mu_\ell > E_\ell^+$. 
This condition involves two direct restrictions. 
One is that the pion momentum must be less than some critical momentum,
\begin{equation}
|\boldsymbol{p}|<\frac{m_\pi^2+m_\ell^2}{2m_\ell^2} q_F - \frac{m_\pi^2-m_\ell^2}{2m_\ell^2}\mu_\ell 
\equiv p_c,
\label{pc}
\end{equation}
where $q_F=\sqrt{\mu_\ell^2-m_\ell^2}$ is the Fermi momentum of the leptons at zero temperature.
The other relation is the one related with the lepton chemical potential, and it must be greater that a critical chemical potential,
\begin{equation}
\mu_\pi > \frac{m_\pi^2+m_\ell^2}{2m_\pi}\equiv \mu_c.
\end{equation}

If we consider the vacuum pion mass $m_\pi = 139.5$~MeV, the critical lepton chemical potential for muons ($m_\mu = 105.6$~MeV) is $\mu_c= 109.74$~ MeV.

\section{Thermal equilibrium}

As was pointed previously, the interpretation of the decay as the imaginary pole of the retarded green function is related with the time that a system slightly out of equilibrium takes to reach the thermal equilibrium. 
In addition to the pions, we also need to consider another process that involves the same participants \cite{Kuznetsova:2008jt}: leptons recombined with antineutrinos. This process of virtual neutrinos from the heat bath captured by leptons increases the pion population. 
we will explore which are the necessary conditions for a system of pions
and leptons in order to reach thermal equilibrium simultaneously.
This happens if the decay rates of pions and the rate of lepton recombination are of similar magnitude $\Gamma_\ell \sim\Gamma_{\pi^-}$.

The lepton rate is obtained also from the retarded propagator of leptons through weak interaction corrections described by the interaction Lagrangian defined in Eq.~(\ref{L_pi-l}).
the dressed lepton propagator can be written as
\begin{equation}
S_\ell^{\textrm{dr}}=
\frac{\slashed{Q}+m-\Sigma}{Q^2-m_\ell^2-\Pi_+}{\cal P_+}
+
\frac{\slashed{Q}+m-\Sigma}{Q^2-m_\ell^2-\Pi_-}{\cal P_-},
\end{equation}
where $Q=(q_0+\mu_\ell,\bs{q})$, $\Sigma=\slashed{A}(1-\gamma_5)$ is the lepton self-energy through weak interactions,
$ {\cal P}_\pm =\frac{1}{2}(1\pm\gamma_0{\boldsymbol \gamma}\punto \hat{\boldsymbol  q}\gamma_5) $, are the helicity projectors and 
$ \Pi_\pm(q) =2Q\cdot A\pm (Q_0{\boldsymbol A}\punto\hat{\boldsymbol  q}-|{\boldsymbol q}|A_0) $
the mass corrections for each helicity.
Now the procedure is the same as in Sec.~\ref{decay_width} by setting  $q_0\to q_0+i\epsilon$ in the retarded propagator. 
The recombination rates for each lepton helicity is
\begin{equation}
\Gamma_{\pm} = -\frac{1}{E_\ell}\mathrm{Im} ~\Pi_\pm(E_\ell-\mu_\ell+i\epsilon,~{\boldsymbol q}),
\end{equation}
where $E_\ell=\sqrt{\bs{q}^2+m_\ell^2}$ is the lepton energy.
In order to simplify the analysis, we  calculate the average of the rates for the different helicities. 
Defining $\Gamma_\ell = (\Gamma_++\Gamma_-)/2$, after Matsubara summation, the average lepton recombination rate is
\begin{align}
\Gamma_{\ell}  = \bar\Gamma_{\pi\ell} \left(\frac{m_\pi}{2m_\ell}\right)^3
\frac{m_\ell}{E_\ell}
&
\frac{T}{2b_\ell |{\boldsymbol q}|}
\Bigg[ 
\ln\left(\frac{1-e^{-(E_\pi^+ -\mu_\pi)/T}}{1-e^{-(E_\pi^- -\mu_\pi)/T}}\right)\;,
\nonumber \\ &
-\ln\left(
\frac{1+e^{-( \tilde E_{\nu_\ell}^+ +\mu_{\nu_\ell})/T}} {1+e^{-( \tilde E_{\nu_\ell}^- +\mu_{\nu_\ell})/T}}\right) 
\Bigg] \;,
\label{Gamma_ell}
\end{align}
where again the pion decay constant in vacuum $\bar\Gamma_{\pi\ell}$ is present. 
The pion and neutrino energy terms are defined as $E_\pi^\pm  =  (1+b_\ell)E_\ell \pm b_\ell |{\boldsymbol q}| $ and $ \tilde E_{\nu_\ell}^\pm   = b_\ell E_\pi \pm b_\ell  |{\boldsymbol q}|$, respectively, with the constant $b_\ell = (m_\pi^2-m_\ell^2)/2m_\ell^2$.

Now, we need to find a window in the parameter space where pions and leptons reach thermal equilibrium satisfying $\Gamma_\ell\sim\Gamma_{\pi^-}$.
The parameters involved are the neutrino chemical potential, the lepton chemical potential, the temperature, the pion energy and the lepton energy. 
To reduce the number of parameters we consider pions  in the rest frame, where the decay rate is given in Eq. (\ref{Gamma1}).
In the same way, near the Fermi surface, fluctuations of leptons in the degenerated environment are produced.
So we take the lepton energy as the Fermi energy in the decay rate, which leads to
\begin{multline}
\Gamma_{\ell}  = \bar\Gamma_{\pi\ell} \left(\frac{m_\pi}{2m_\ell}\right)^3
\frac{m_\ell}{\mu_\ell}
\frac{T}{2b_\ell\, q_F}
\\ \times
\ln\left(\frac{\sinh [(b_\ell\mu_\ell+\mu_{\nu_\ell})/T]+\sinh[b_\ell q_F/T]}{\sinh [(b_\ell\mu_\ell+\mu_{\nu_\ell})/T]-\sinh[b_\ell q_F/T]}\right).
\end{multline}

\begin{figure}[t]
	\centerline{\includegraphics[scale=.7]{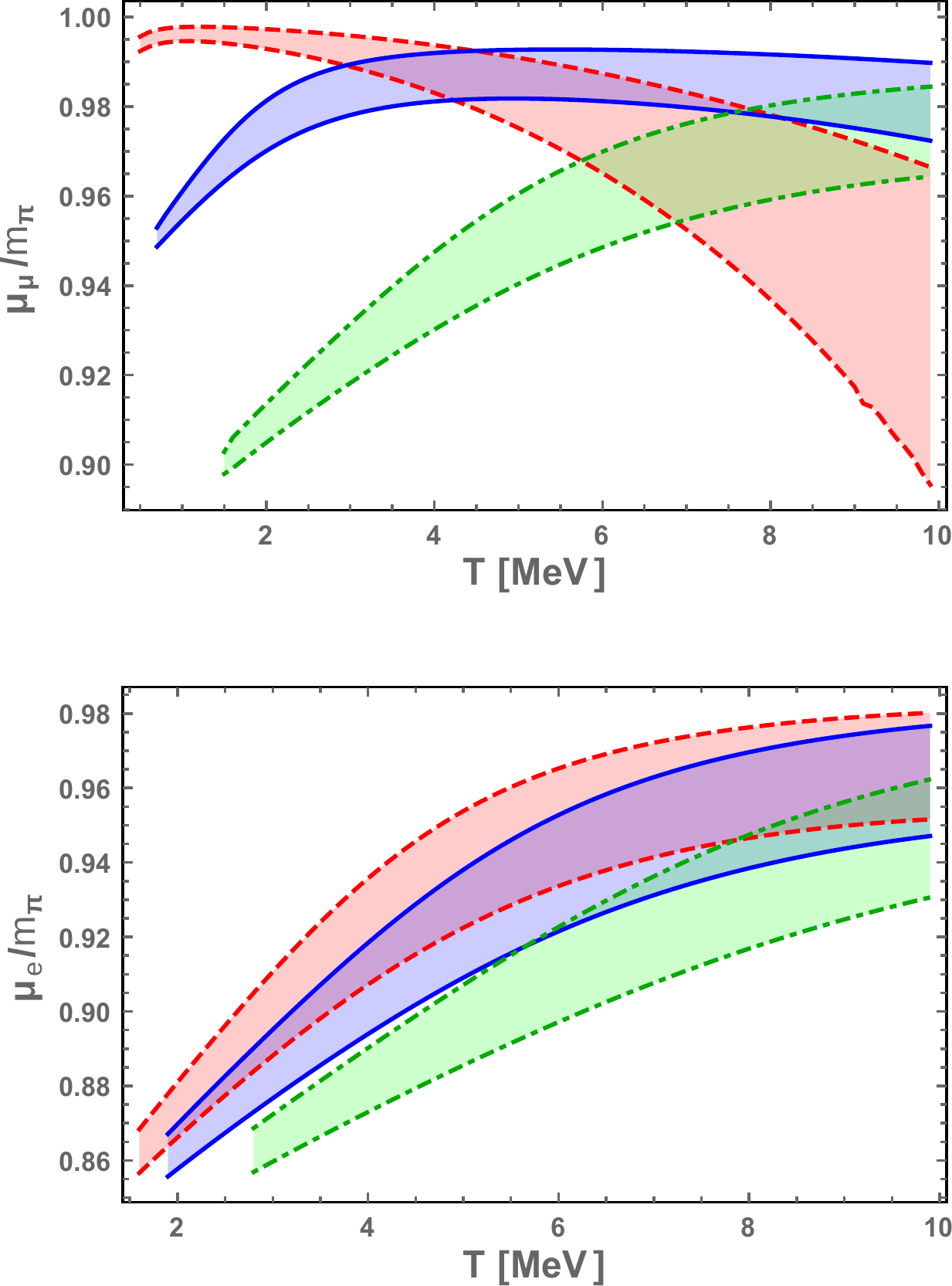}}
	\caption{Chemical potential and temperature values where pion-lepton chemical equilibrium is favorable.
The band widths correspond to a half order of magnitude difference between the widths: $0.5 < \Gamma_{\pi^-}/\Gamma_{\ell} < 1.5$.
The colors and lines refer to $m_\pi =140$~MeV (solid blue),  $m_\pi =115$~MeV (dashed red) and $m_\pi =200$~MeV (dot-dashed green).}
\label{FIG-widths}
\end{figure}

Figure~\ref{FIG-widths} shows the leptonic chemical potential as a function of temperature necessary to have simultaneous thermal equilibrium for lepton pion system, considering pions in rest frame and leptons at the Fermi surface.
The band considers the possibility to fluctuate a bit, with $\frac{1}{2}\Gamma_\ell < \Gamma_{\pi^-}<\frac{3}{2}\Gamma_\ell$. 
For simplicity we  consider $\mu_{\nu_\ell}=0$, which means that all real neutrinos escape from the star once created. 
This is a usual approximation in neutron stars, but not necessarily valid for protoneutron stars
\cite{Pons:2000xf}.

From the two plots, for the pion muon process (upper plot) and for pion electron process (lower plot), at least the region for $T>1$~ MeV considers values of the lepton chemical potential in the metastability region, $\mu_\ell > \mu_c$. 
For lower temperatures this is still valid for temperatures less than 1~ keV, but we are interested in this region because it will be the region of higher emissivity as we will show next

\section{Emissivity}

\begin{figure}[t]
	\centerline{\includegraphics[scale=.7]{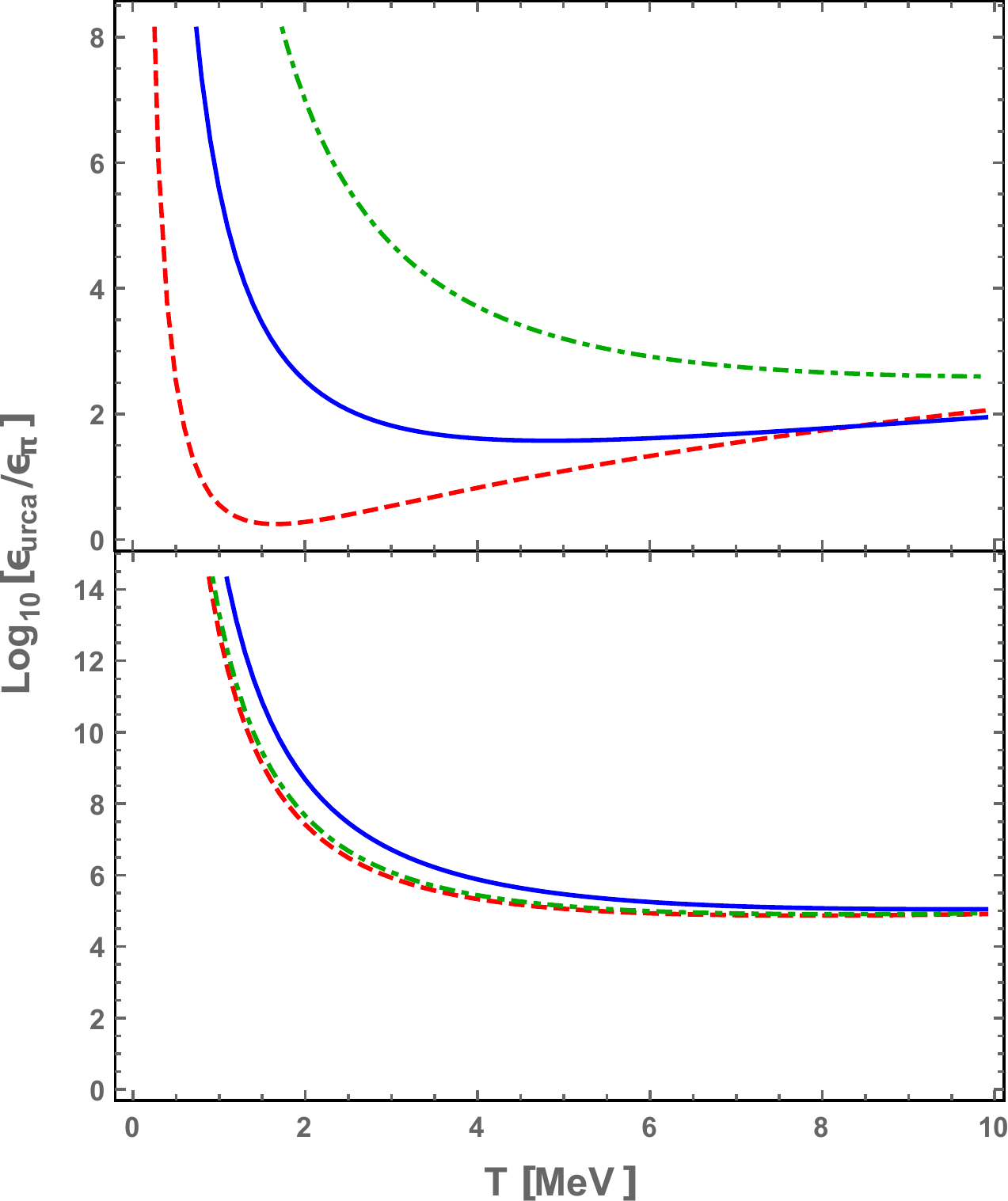}}
	\caption{Comparison between emissivities produced by the URCA process and pion decaying into muons (upper panel) and electrons (lower pannel). 
We use the the chemical potential value $\mu_\mu=0.95 m_\pi$ and the pion masses $m_\pi$ =115 (dashed red), 140 (solid blue), and 200 MeV (dot-dashed green). }
    \label{FIG-emissivity}
\end{figure}

One important consequence of charged pion metastability is the neutrino emission in this soft decay process. 
The most efficient mechanisms in neutron stars are the URCA and modified URCA process. 
The emissivity is related with the cooling process, which is the main insight to speculate of different composition of matter in the inner core of a neutron star related with the cooling process.
Considering leptonic pion decay, the neutrino emissivity is defined as the energy of the ejected antineutrino multiplied by the probability of finding a pion, $n_{\pi^-}$, in the thermal bath and the probability of finding a hole below the Fermi level in order to put a lepton, $1-n_\ell$, all this integrated in phase space
\begin{multline}
\epsilon_{\pi} =  
\int \bar{d}p\,\bar{d}q\,\bar{d}k\,
\sum_{\text{spin}}|{\cal M}|^2\, k_0\, n_B(p_0)\, [1-n_F(q_0)]\\
(2\pi)^4\delta^4(p-q-k),
\end{multline}
where the phase space measure for pions is
\begin{equation}
\bar dp = \frac{d^4 p}{(2\pi)^3}\theta(p_0+\mu_\pi)\delta^4((p_0+\mu_\pi)^2-E_\pi^2)\;,
\end{equation}
and an equivalent expression for the leptons ($q$) and neutrinos (k) with their respective chemical potentials.
The probability amplitude for pions going into leptons and antineutrinos is defined as 
\begin{equation}
\langle \ell\,\bar\nu| \!\int\!\! d^4x\, {\cal L}_{\pi\ell}\, |\pi^-\rangle
=i{\cal M}(2\pi)^4\delta^4(p-q-k).
\end{equation}
At chemical equilibrium, the neutrino emissivity from charged pions decay is then
\begin{multline}
\epsilon_{\pi} = \frac{\bar\Gamma_{\pi\ell}\, m_\pi T^2}{2\pi^2 a_\ell}
\int_{m_\pi}^\infty dE_\pi \, n_B(E_\pi-\mu_\pi) 
\\
\bigg[
\frac{E_\pi-E_\ell}{T}\,\ln\!\left(1+e^{(E_\ell-\mu_\ell)/T}\right)
\\
\left.
-{\rm Li}_2\!\left(-e^{(E_\ell-\mu_\ell)/T}\right)
\bigg]\right|_{E_\ell^-}^{E_\ell^+},
\label{emissivity_full}
\end{multline}
where the integral variable $E_\ell$ is evaluated in $E_\ell^\pm$, defined after Eq. (\ref{Gamma_pi}). 
This result is compared with the URCA emissivity \cite{Yakovlev:2000jp}, as is shown in FIG.~ \ref{FIG-emissivity}, considering the ratio of URCA process emissivity and the emissivity produced by pions decaying into leptons, as a function of temperature, for different values of the pion mass.
Here, we assume that all the neurinos escape from the system by setting $\mu_{\nu_\ell}=0$. 
For the pion decay constant and also for nucleon masses in the URCA emissivity, we use the Brown-Rho scaling \citep{Brown:1991kk}, with a scaling factor 0.8.
Here we show the region where the emissivity from pion decay reaches its maximum value compared with the URCA emissivity.
Note that for low pion mass and temperatures of the order of 1~MeV, the neutrino emissivity from pions decaying into muons can be of the same order of magnitude than the URCA emissivity.
This is the region where, in the case of emissivity, consequences of metastable pions are more appreciable.

The difference in the comparison of neuron emissivity from metastable pions, instead of normal pions, can be more appreciable by approximating  EC. (\ref{emissivity_full}) in a low temperature expansion. 
This is possible since the temperature values we are considering are much smaller than the pion mass.
When the critical momentum defined in Eq(\ref{pc}) is much larger than the temperature $p_c\gg T$, which is the case in the region of temperature we show in Fig.~\ref{FIG-emissivity} 
the emissivity is proportional to $T^2$ times an exponential suppression term
\begin{equation}
\epsilon_\pi \approx
\bar\Gamma_{\pi\ell} \,m_\pi^4\,
g(\mu_\ell)\,\left(\frac{T}{2\pi m_\pi}\right)^{2}
 e^{-(E_c-\mu_\pi)/T},
\label{emissivity-approx}
\end{equation}
with 
\begin{multline}
g(\mu_\ell)=
\frac{2q_F}{b_\ell m_\pi}+\frac{2a_\ell m_\pi}{p_c}+\frac{2p_c}{a_\ell m_\pi}\frac{q_F-p_c}{q_F+p_c}, 
\end{multline}
with the critical energy defined as $E_c=\sqrt{p_c^2+m_\pi^2}$, 
and all the other terms defined after EQ. (\ref{Gamma_pi}), (\ref{pc}) and (\ref{Gamma_ell}).
The result in Eq. (\ref{emissivity-approx}) shows a different behavior compared with the emissivity of a simple pion gas \cite{Jaikumar:2002vg}, where $\epsilon\sim T^{3/2}e^{-(m_\pi-\mu_\pi)/T}$, valid for $m_\pi-\mu_\pi \gg T$.

The result of Eq. \ref{emissivity-approx} is a direct example of how different is the behavior of a pions with the Pauli blocking restriction to their consequent particle products after their weak decay.

\section{conclusions}

Here we investigate the possibility of a low decaying state of charged pions in a dense of degenerated lepton system. 
The study considers neutron star or protoneutron star environments where pions and leptons are in chemical equilibrium.

It worth to mention that all the results obtained in this work, are a direct consequence that appears naturally. In other words, we do not introduce any extra ingredient in order to force the system to behave differently.

We can show that it is possible to find such state of matter, metastable charged pions in thermal equilibrium with a low decaying process inside compact stars.
This condition has a direct impact in the emissivity of neutrinos from the pion decay process with a behavior of $\epsilon\sim T^2e^{-(E_c-\mu_\pi)/T}$ instead of the usual pion gas emissivity  $\epsilon\sim T^{3/2}e^{-(m_\pi-\mu_\pi)/T}$ which is commonly used.

The conditions for thermal equilibrium are basically satisfied in the range of interest, where the lepton chemical potential is higher than the critical chemical potential needed for metastability, and the temperature where emissivity is maximal, between $T\approx 1-10$~MeV. 

Although the emissivity is low compared with URCA process, and even much less than the modified URCA process, there are many other factors that can change this estimations. For the moment, the possibility of this source of neutrino emission be of the  same order than the URCA neutrino emission, for low pion mass case at $T\sim 1$~MeV, could be a good signal.

For now, this state of matter  doesn't seem to be a very significant contribution, other than being another source of neutrinos. 
However, many estimations that considers pion in unstable state can be revisited. 
The effects of considering metastable pions and not unstable pions in particle populations may have a direct impact.


\vspace{-.1cm}

\section*{Acknowledgments}


 The authors acknowledge the support of 
\fundingAgency{FONDECYT} (Chile) under Grants No. 
    \fundingNumber{1130056, 1150471 and 1150847}; 
\fundingAgency{ConicytPIA/Basal} (Chile) Grant No. 
    \fundingNumber{FB0821};
\fundingAgency{CIC-UMSNH} (M\'exico) Grant No. 
    \fundingNumber{4.22};
\fundingAgency{CONACyT} (M\'exico) Grant No.
    \fundingNumber{256494}

\vspace{-.1cm}


\end{document}